\documentstyle[epsf]{article}
\begin{document}


\title{Recent Physics At COSY--A Review}

\author{H. Machner\\Institut f\"{u}r Kernphysik,\\
Forschungszentrum J\"{u}lich, 52425 J\"{u}lich, Germany}

\maketitle


\begin{abstract}
The COSY accelerator in J\"{u}lich is presented together with its
internal and external detectors. The physics program performed
recently is discussed with emphasis on strangeness physics and
precision experiments. \end{abstract}
\section{Introduction}
The accelerator complex at J\"{u}lich consists of an isochronous
cyclotron as injector and a strong focusing synchrotron. Injection
is performed by stripping of negative ions (hydrogen and deuterium)
at an energy of 40 MeV*A. They can be vector polarised (protons and
deuterons) and tensor polarised (deuterons). The synchrotron COSY is
equipped with electron cooling at injection energy and stochastic
cooling at higher energies. It provides beams up to $3700$ MeV/c
momentum. COSY can be used as a storage ring to supply internal
experiments with beam. The beam can also be stochastically extracted
within time bins ranging from 10 s to several minutes to external
experiments. The emittance of the extracted cooled beam is only
$\epsilon=0.4 \pi$ mm mrad. This allows excellent close to target
tracking. Hence a large fraction of the experimental program is
devoted to meson production close to threshold.
\begin{figure}[h]
\begin{center}
\epsfxsize=13cm \centerline{\epsfbox{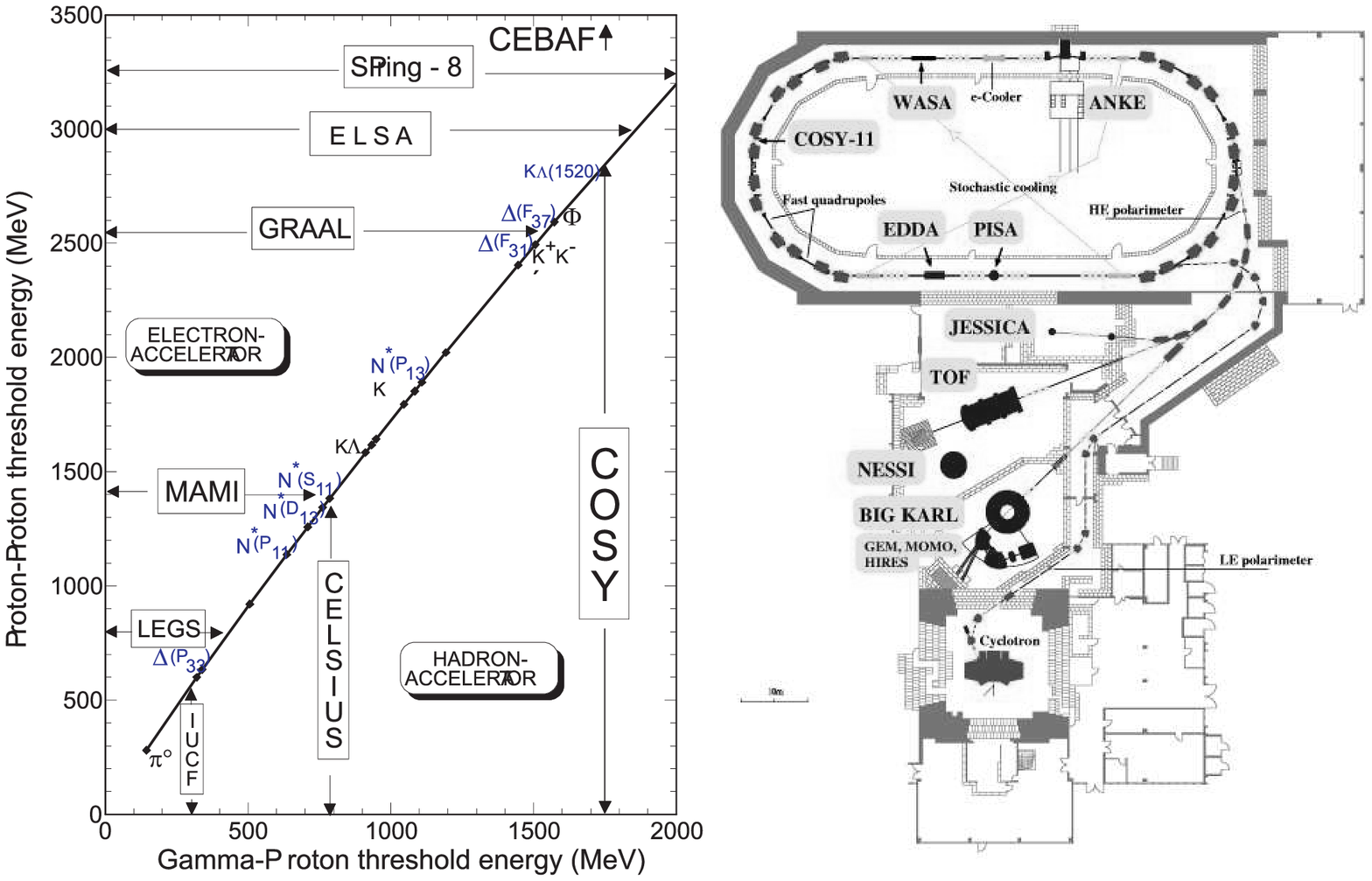}}
\caption{\emph{Left}: Energy dependence of photon beams and proton
beams in order to produce indicated mesons or nucleon resonances
indicated in the figure. The maximal energy of some accelerators are
indicated as well. \emph{Right}: Floor plan of COSY. Positions of
internal experiments as well as those of external experiments are
shown.} \label{Energy_Comp}
\end{center}
\end{figure}
In Fig. \ref{Energy_Comp} is the maximal energy of COSY compared to
other hadron accelerators. Some mesons, which can be produced in a
$pp$-collision in the COSY range, are shown together with some
selected nucleon resonances. Also shown is the equivalent photon
energy to produce the same final systems and maximal energies of
photon facilities are given.

Here we will concentrate on hadron physics thus leaving out
detectors built for different purposes like PISA, NESSI and JESSICA
(see Fig. \ref{Energy_Comp}). The detectors of interest are COSY-11,
ANKE and EDDA internally and TOF and BIG KARL externally. The
physics at EDDA, proton-proton scattering of unpolarised on
unpolarised, polarised on unpolarised and polarised on polarised
\cite{EDDA04} is terminated and is therefore neglected here. COSY-11
and ANKE are magnetic detectors. The former \cite{COSY11} employs an
accelerator dipole magnet while the latter \cite{ANKE} is a chicane
consisting of three dipoles with the middle one as analysing magnet.
TOF \cite{TOF} is a a huge vacuum vessel with several layers if
scintillators. Time of flight is measured between start detectors in
the target area and the scintillators. The target area detectors are
especially suited for the identification of delayed decays and TOF
is thus a geometry detector. BIG KARL \cite{Drochner98,Bojowald02}
is a focusing magnetic spectrograph of the 3Q2D-type. Particle
tracks are measured in the focal plane area with packs od MWDC's
followed by scintillator hodoscopes allowing for a time of flight
path of 3.5 m. Additional detectors exist. MOMO \cite{Bellemann99}
measures the emission vertex of charged particles. The Germanium
Wall \cite{Betigeri99} is a stack of four annular germanium diodes
being position sensitive. It acts as a recoil spectrometer.

\section{Strangeness Physics}
One strong item in COSY physics is the study of strangeness
production in various processes in $pp$, $pd$ and $pA$ interactions.
Here we have to concentrate on  a few of these reactions.

\begin{figure}[h]
\begin{center}
\epsfxsize=8cm \centerline{\epsfbox{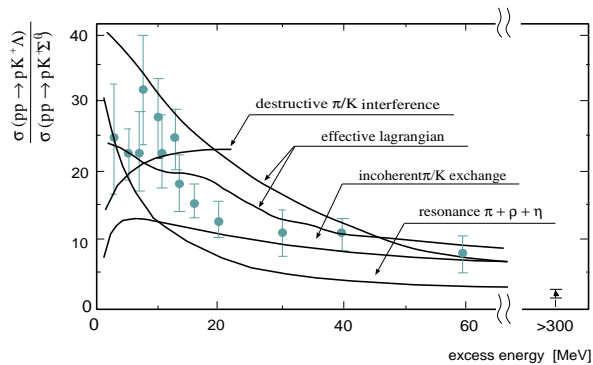}}
\caption{Ratio of the cross sections for the indicated associated
strangeness production. The curves are model calculations discussed
in the text.} \label{Associated}
\end{center}
\end{figure}

The $pp\to pK\Lambda(\Sigma)$ reactions, associated strangeness
production were measured by COSY-11 \cite{Bal98,Sew99,Kowina04} and
TOF \cite{Marcello01}.  Fig. \ref{Associated} shows the ratio
$\sigma(pK^+\Lambda)/\sigma(pK^+\Sigma^0)$ as function of the excess
energy.
The ratio rises strongly to threshold. This unexpected behaviour is
studied within several models, including pion and kaon exchange
added coherently with destructive interference \cite{Gasparian00} or
incoherently~\cite{Sibirtsev99}, the excitation of nucleon
resonances~\cite{Shyam01,Shyam04} (labeled effective Lagrangian),
resonances with heavy meson exchange ($\pi,\rho,\eta$)
\cite{Sibirtsev00} and heavy meson exchange ($\rho$, $\omega$ and
$K^*$)~\cite{Shyam01,Shyam04}. The corresponding curves are also
shown in the figure. All models show a decrease of the ratio with
increasing excitation energy but none of them accounts for all data.

The associated strangeness production is also a useful tool to study
the nucleon-hyperon interaction via FSI. At present a high
resolution study of this interaction runs at BIG KARL (see
contribution by Roy et al.). The measurement of Dalitz diagrams
enables even the investigation of the importance of intermediate
$N^*$ excitation.

Connected with the associated strangeness production is the question
whether pentaquarks exist. Most of the experimental searches were
performed with electromagnetic probes on the neutron, which, of
course, is embedded in a nucleus. A cleaner environment is the $pp$
interaction.
\begin{figure}[h]
\begin{center}
\epsfxsize=8cm \centerline{\epsfbox{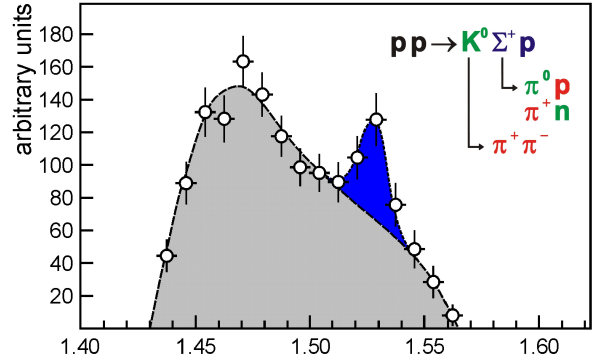}}
\caption{Evidence for a pentaquark produced in $pp$ collision.}
\label{Pentaquark}
\end{center}
\end{figure}
The reaction studied with TOF is the
\begin{equation}
pp\to K^0\Sigma^+p
\end{equation}
reaction. The $K^0$ is identified via its decay into two pions and
the $\Sigma$ via its delayed decay. The data from Ref.
\cite{Abdel_Bary04} are shown in Fig. \ref{Pentaquark}. There is
evidence on a $4\sigma$ level for the production of a pentaquark
\begin{equation}
pp\to \Theta^+\Sigma^+
\end{equation}
with a subsequent decay of the $\Theta^+$ into $K^0$ and $p$. For
the enormous body of papers related to pentaquark we refer to a
review given by Stancu \cite{Stancu05}.

Another interesting reaction is
\begin{equation}
pp\to d K^+\bar K^0.
\end{equation}
On order to reach the threshold of this reaction the maximal energy
of COSY had to be lifted above its design value of 2.5 GeV. The data
were taken at an energy of 2.65 GeV \cite{Kleber03}. The analysis of
the data \cite{Grishina04} resulted in a dominance of the channel
\begin{equation}
pp\to d a_0^+
\end{equation}
with a subsequent decay $a_0^+\to K^+\bar K^0$.

At the BIG KARL spectrograph the reaction
\begin{equation}
pd\to {^3He}K^+K^-
\end{equation}
was studied with the MOMO vertex wall.
\begin{figure}
\epsfxsize=13 cm
\centerline{\epsfbox{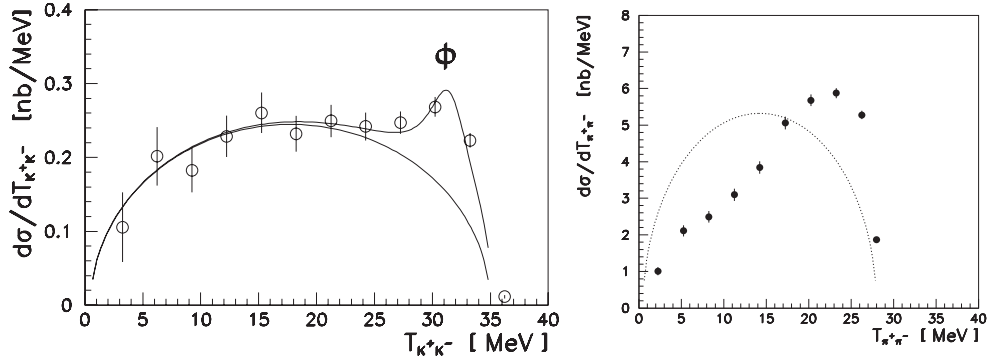}}
\caption{Left: Energy spectrum of the two kaons at a maximal excess
energy of 35 MeV. Right: Same as left but for two pions for a
maximal excess energy of 28 MeV.} \label{KK_pipi}
\end{figure}
The interest in this reaction stems from the surprising behaviour of
two pions in the
\begin{equation}
pd\to {^3He}\pi^+\pi^-
\end{equation}
reaction \cite{Bellemann99}. The latter reaction showed a p-wave
between the two pions even close to threshold. In Fig. \ref{KK_pipi}
the energy spectrum of two kaons for a maximal energy of 35 MeV are
shown. Besides a smooth continuum the production of $\phi$-mesons is
visible. The energy distribution follows phase space, hence it is
s-wave. The same conclusion  holds for the $KK$--${^3He}$ system.
Also the excitation function for the $\phi$ production is an accord
with the assumption of s-wave. To summarise: the $KK$--${^3He}$
behaves as expected while the $\pi\pi$--${^3He}$ system shows an
unexpected behaviour. In order to proof the findings for this system
the experiment was repeated but with inverse kinematics. The
advantage of doing so is a smaller number of settings of the
spectrograph. A result of this measurement is also shown in Fig.
\ref{KK_pipi}. It supports the previous findings.

One aspect of strangeness physics is the $s\bar s$ content in the
nucleon. This is connected to a violation of the OZI-rule in the
ratio
\begin{equation}\label{eqn:OZI}
R=\frac{\sigma(pp\to pp\phi)}{\sigma(pp\to pp\omega)}.
\end{equation}
These two mesons have almost ideal quark mixing and hence the
$\omega$ has negligible $s\bar s$ content while the $\phi$ is an
almost pure $s\bar s$ state (see previous reaction). TOF measured
$\omega$ production at exactly the same excess energy as previous
$\phi$ production, thus allowing the deduction of $R$ as function of
excess energy. This yields $R=(3\pm 1)\times 10^{-2}$ while the
OZI-rule predicts $R=4\times 10^{-3}$. This may point to a serious
content of  $s\bar s$ pairs in the nucleon.

\section{$\eta$-meson Physics, Symmetries}

There is a wealth of data of light meson production measured at COSY
in the threshold region. The data were taken mainly by the COSY-11
and GEM collaborations for the nucleon-nucleon channel
\cite{Machner99,Moskal_Review,Betigeri02}. Here we will concentrate
on $\eta$ and $\eta '$ production.
\begin{figure}[h]
\begin{center}
\epsfxsize=6 cm \centerline{\epsfbox{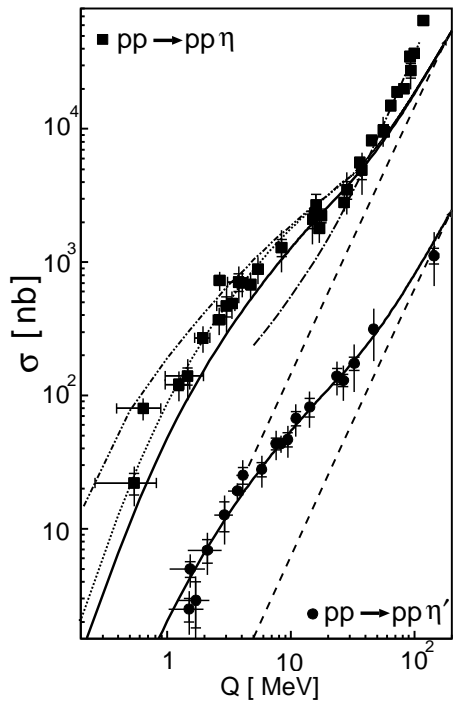}}
\caption{Total cross sections for the indicated reactions as
function of the excess energy $Q$. For an explanation of the
different curves see text.} \label{Eta_Exfu}
\end{center}
\end{figure}
First, we discuss the production in $pp$ interactions. Fig.
\ref{Eta_Exfu} shows the total cross sections \cite{Moskal05} and
references therein as function of the excess energy $Q$. The dashed
lines indicate a phase space integral normalised arbitrarily. The
solid lines show the phase space distribution with inclusion of the
$^1S_0$ proton-proton strong and Coulomb interactions. In case of
the $pp\to pp\eta$ reaction the solid line was fitted to the data in
the excess energy range between 15 and $40\,\mbox{MeV}$. Additional
inclusion of the proton-$\eta$ interaction is indicated by the
dotted line. The scattering length of $a_{p\eta} = 0.7\,\mbox{fm} +
i\,0.4\,\mbox{fm}$ and the effective range parameter $b_{p\eta} =
-1.50\,\mbox{fm} -i\,0.24\,\mbox{fm}$ have been chosen arbitrarily.
The dashed-dotted line represents the energy dependence taking into
account the contribution from the $^3P_{0}\to ^1\!\!\!S_{0}s$,
$^1S_{0}\to ^3\!\!\!P_{0}s$ and $^1D_2\to ^3\!\!\!P_2 s$
transitions. Preliminary results for the $^3P_{0}\to ^1\!\!\!S_{0}s$
transition with full treatment of three-body effects are shown as a
dashed-double-dotted line. The absolute scale of
dashed-double-dotted line was arbitrary fitted to demonstrate the
energy dependence only. To summarise: in $\eta '$ production the FSI
between the two protons is of importance. In the case of $\eta$
production also the FSI between the $\eta $ and the two protons
matters. This rather strong interaction has led to the speculation
wether bound or quasi-bound $\eta$-nucleus systems may exist. This
will be discussed in other talks at this meeting.

The PDG \cite{PDG} reported up to 2002 a rest mass of the
$\eta$-meson of $(547.30\pm 0.12)$ MeV/c$^2$. Then a new measurement
by the NA48 group at CERN reported a larger value \cite{Lai02} with
a very small uncertainty: $(547.843\pm 0.030\mbox{ stat. }\pm
0.041\mbox{ syst.})$ MeV/c$^2$. This value does not within error
bars overlap with the previous results. This measurement made use of
the decay $\eta\to 3\pi^0$. GEM performed an experiment with a
different technique proof this result with uncertainties within the
same order of magnitude. The idea is a self calibrating experiment
making use of the following three reactions:
\begin{eqnarray}
p+d&\to {^3H}+\pi^+, \label{eqn:rea1} \\
p+d&\to \pi^++{^3H}, \label{eqn:rea2}\\
p+d&\to {^3He}+\eta. \label{eqn:rea3}
\end{eqnarray}
At a beam momentum around $1640$ MeV/c the ${^3H}$ ions being
emitted forward, the $\pi^+$'s being emitted backward in the centre
of mass system as well as the ${^3He}$ ions being emitted backward
are within the acceptance of the magnetic spectrograph. Reactions
\ref{eqn:rea1} and \ref{eqn:rea2} are in principe used to calibrate
the detector and the accelerator. Reaction \ref{eqn:rea3} is then
used to determine via the missing mass technique the $\eta$ mass. It
was found to be
\begin{equation}
m(\eta)=547.311 \pm 0.028\mbox{ stat. }\pm 0.032\mbox{ syst.})
\mbox{MeV/c}^2.
\end{equation}
Details of the experiment can be found in Ref. \cite{Abdel-Bary05}.
\begin{figure}[h]
\begin{center}
\epsfxsize=6 cm \centerline{\epsfbox{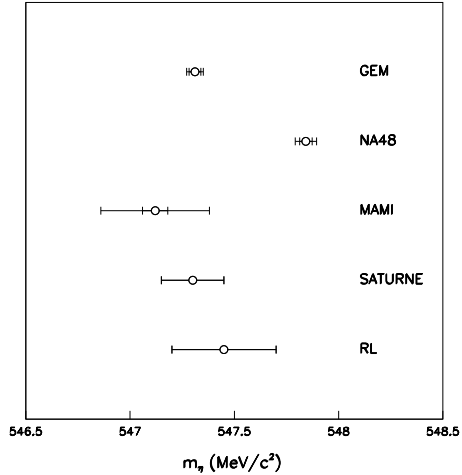}} \caption{The
$\eta$-meson rest mass as published by different experiments in the
order of their publication. The references given can be found in the
particle data book.} \label{Eta_Mass}
\end{center}
\end{figure}
This new value is compared with previous data in Fig.
\ref{Eta_Mass}. It shows excellent agreement with the previously
published values except the one from Ref. \cite{Lai02}.

Now we will proceed to $pd\to {^3A}X $ reactions with $A=H$ or $He$
and $X=\pi,\eta$. Here we will start with pion production. A series
of angular distributions was measured by GEM, in order to study
possible violations of isospin symmetry
\cite{Betigeri01,Abdel-Samad03}. Two components were found: one
depending strongly on the emission angle and thus on the momentum
transfer, another one which is almost isotropic. In the first
component a violation of isospin symmetry was found. The data also
serve to investigate the production mechanism. This has to be
understood before more complicated reactions as multi-pion
production or $\eta$ production can be understood.
\begin{figure}[h]
\begin{center}
\epsfxsize=8cm \centerline{\epsfbox{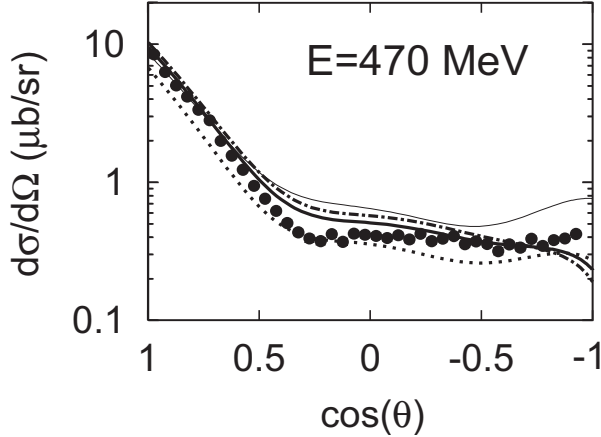}}
\caption{Angular distribution of the $p+d\to {^3He}+\pi^0$ reaction
at the indicated beam energy. Solid, dash-dotted and dotted lines
denote the calculations for Bonn-B, Bonn-A, and  Paris $NN$
potentials, respectively,  while thin solid line corresponds to the
plane-wave calculation (without the $pd$ ISI).} \label{pd2pihe3}
\end{center}
\end{figure}
A calculation which goes beyond the Locher-Weber type calculations
\cite{Betigeri01,Locher74} was performed recently by Canton and
Levchuk \cite{Canton05}. They treated the three nucleon wave
functions in the ISI within the AGS formalism. Their results are
compared with GEM data in Fig. \ref{pd2pihe3}. One can see that the
isotropic component strongly depends on ISI. The Paris $NN$
potential seems to slightly underestimate the experiment.

\begin{figure}
\epsfxsize=13cm \centerline{\epsfbox{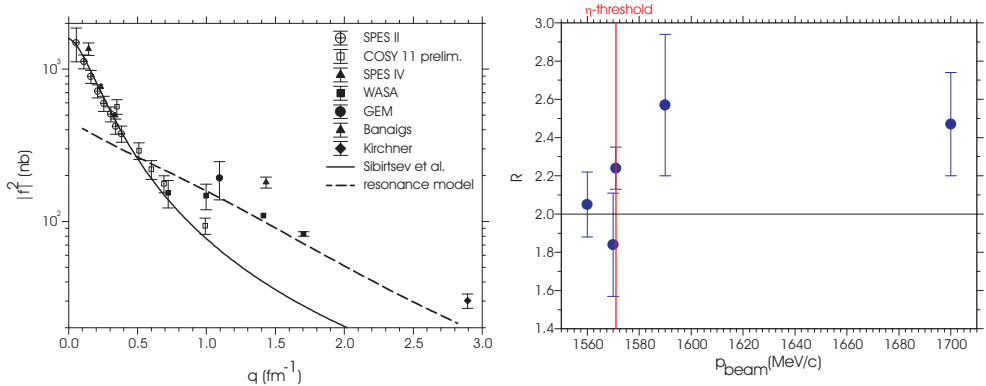}}
\caption{Left: Excitation function for the $pd\to {^3He}\eta$
reaction. The references can be found in 
Right:
Excitation functions for the ratio of the two pion production
reactions at maximal momentum transfer (zero degree in the lab.
system). The $\eta$-production threshold is indicated as line.}
\label{iso_threshold}
\end{figure}

We now proceed to the $p+d\to {^3He}+\eta$ reaction which is of
great interest, since the $\eta$-nucleus interaction should be more
attractive than in the $pp$ case. From the differential cross
sections one can deduce a matrix element $f$ as
\begin{equation}\label{equ:Matrix}
|f\left( \theta  \right)|^2  = \frac{{k^2 }}{{q^2 }}\frac{{d\sigma
}}{{d\Omega }}\left( \theta  \right) = |f_p |^2 |T\left( q
\right)|^2.
\end{equation}
The right hand side is valid for $s$-wave production with FSI in the
exit channel. With $f_p$ the production matrix element and with $T$
the FSI matrix element are denoted. Willis et al. \cite{Willis97}
analysed near threshold data in this way. Later, Sibirtsev et al.
\cite{Sibirtsev03} extended the range to the then available data.
Their value for the $\eta-{^3He}$ scattering length is
$a(\eta-{^3He})=(-4.3{\pm}0.3)+i(0.5{\pm}0.5)$. The resulting angle
integrated matrix element is shown in the left part of Fig.
\ref{iso_threshold} as function of the of the $\eta$ momentum $q$.
This is in agreement with model calculations for instance those of
Khemchandani et al. \cite{Khemchandani_02}. They calculated the
cross section in a two-step model. In a first step a pion is
produced in a nucleon-nucleon interaction while in the second step
this pion is scattered at the spectator nucleon $\pi+N\to \eta +N$
with eventual fusion of the three nucleons. The agreement between
data and \ref{equ:Matrix} is nice up to $q_\eta \approx 1$ fm$^{-1}$
(the data from COSY 11 are still preliminary). Also shown in the
figure is the matrix element of a resonance model \cite{Betigeri00}
without FSI enhancement. Therefore it underestimates the data close
to threshold. There are ambiguities between the data. Angular
distributions close to threshold are isotropic as expected. At
higher energies the data from Ref.'s
\cite{Betigeri00,Banaigs73,Kirchner93} show a strong forward peaking
while the data from \cite{Bilger02} peak at $\cos\theta_\eta\approx
0.5$. This could be reproduced by the calculations performed by
Stenmark \cite{Stenmark03}. However, this approach is based on the
ad hoc assumption of the intermediate pion to be limited into a
narrow forward cone, as was pointed out in Ref.
\cite{Khemchandani_03}.

The GEM collaboration studied isospin symmetry breaking by comparing
neutral and charged pion production in $pp\to d\pi^+$ and $np\to
d\pi^0$ \cite{Drochner98} reactions, and $pd\to {^3H}\pi^+$ and
$pd\to {^3He}\pi^0$ reactions \cite{Abdel-Samad03}. For the latter
reactions it was found that the angular distribution of the matrix
elements consists of two parts. an exponential part showing scaling
which is attributed to a one step reaction. This part shows isospin
symmetry breaking. The second component is isotropic and is related
to two step processes. It does not show isospin symmetry breaking.
The origin of isospin symmetry breaking is in addition to the
Coulomb force a difference in the masses of the up and down quark.
It was suggested \cite{Magiera00} to study the $pd\to {^3He}\pi^0$
reaction at maximal momentum transfer around the $\eta$-production
threshold. This channel should be sensitive to $\pi^0$-$\eta$ mixing
with the mixing angle being dependent on the different quark masses.
On the contrary the $pd\to {^3H}\pi^+$ reaction should not show such
an interference effect. This was indeed found in an experiment
\cite{Abdel-Bary03} and the ratio of both reactions is shown in Fig.
\ref{iso_threshold}. Baru et al. \cite{Baru03} claimed this effect
to be most probably to FSI between the ${^3He}$ and the virtual
$\eta$. However, if the data are analysed on terms of the model from
Ref. \cite{Magiera00} the mixing angle results into $\theta=0.006\pm
0.005$. Green and Wycech used a K-matrix formalism and derived
$\theta= 0.010\pm 0.005$. From this formalism a rather large
$\eta$-nucleon scattering length is extracted making a bound
$\eta$-nucleus very likely. The search for such a system is in
progress (see contributions by Jha and Chatterjee). Also the
question of $\pi^0$-$\eta$ mixing will be further studied via
isospin forbidden decays of $\eta$ and $\eta '$ mesons with WASA at
COSY \cite{WASA}.

\section{Acknowledgments}

I am grateful to the members of GEM for their collaboration. The
contributions and discussions with Dres. M. B\"{u}scher, A. Gillitzer,
R. Jahn, A. Khoukaz, P. Moskal and F. Rathmann are gratefully
acknowledged.


\begin{thebibliography}{10}

\bibitem{EDDA04}
F.~Bauer et~al.: \emph{arXive}, nucl-ex, 0412014 (2004).

\bibitem{COSY11}
S.~Brauksiepe at~al.: \emph{Nucl. Instruments and Meth. in Phys.
Res.}, A 376,
  397 (1996).

\bibitem{ANKE}
S.~Barsov et~al.: \emph{Nucl. Instruments and Meth. in Phys. Res.},
A 462, 364
  (2001).

\bibitem{TOF}
M.~Dahmen et~al.: \emph{Nucl. Instruments and Meth. in Phys. Res.},
A 348, 97
  (1994).

\bibitem{Drochner98}
M.~Drochner, J.~Ernst, S.~F\"{o}rtsch, L.~Freindl, D.~Frekers,
W.~Garske,
  K.~Grewer, S.~Igel, R.~Jahn, L.~Jarczyk, G.~Kemmerling, K.~Kilian,
  S.~Kliczewski, W.~Klimala, D.~Kolev, T.~Kutsarova, G.~Lippert, H.~Machner,
  R.~Maier, C.~Nake, B.~Razen, P.~Von Rossen, B.J. Roy, K.~Scho, R.~Siudak,
  J.~Smyrski, A.~Strzalkowski, R.~Tsenov, P.A. Zolnierczuk, K.~Zwoll:
  \emph{Nucl. Phys.}, A 643, 55 (1998).

\bibitem{Bojowald02}
H.~Bojowald et~al.: \emph{Nucl. Instruments and Meth. in Phys.
Res.}, A 487,
  314 (2002).

\bibitem{Bellemann99}
F.~Bellemann et~al. (COSY-MOMO~Collaboration): \emph{Phys. Rev.}, C
60, 061002
  (1999).

\bibitem{Betigeri99}
M.~Betigeri et~al.: \emph{Nucl. Instruments and Meth. in Phys.
Res.}, A 421,
  447 (1999).

\bibitem{Bal98}
J.T.~Balewski et~al.: \emph{Phys. Lett.}, B 420, 211 (1998).

\bibitem{Sew99}
S.~Sewerin et~al.: \emph{Phys. Rev. Lett.}, 83, 682 (1999).

\bibitem{Kowina04}
P.~Kowina et~al.: \emph{The European Physical Journal}, A 22, 293
(2004).

\bibitem{Marcello01}
S.~Marcello et~al.: \emph{Nucl. Phys.}, A 691, 344c (2001).

\bibitem{Gasparian00}
A.~Gasparian et~al.: \emph{Phys. Lett.}, B 480, 273 (2000).

\bibitem{Sibirtsev99}
A.~Sibirtsev et~al.: \emph{Nucl. Phys.}, A 646, 427 (1999).

\bibitem{Shyam01}
R.~Shyam et~al.: \emph{Phys. Rev.}, C 63, 022202 (2001).

\bibitem{Shyam04}
S.~Shyam: \emph{arXive}, hep-ph, 0406297 (2004).

\bibitem{Sibirtsev00}
A.~Sibirtsev et~al.: \emph{arXive}, nucl-th, 0004022 v2 (2000).

\bibitem{Abdel_Bary04}
M.~Abdel~Bary et~al. (COSY-TOF~Collaboration): \emph{Phys. Lett.}, B
595, 127
  (2004).

\bibitem{Stancu05}
Fl. Stancu: \emph{Int. J. Mod. Phys.}, A 20, 209 (2005).

\bibitem{Kleber03}
V.~Kleber et~al.: \emph{Phys. Rev. Lett.}, 91, 172304 (2003).

\bibitem{Grishina04}
M.~B\"{u}scher W.~Cassing V.~Yu.~Grishina, L. A.~Kondratyuk: A 21,
507 (2004).

\bibitem{Machner99}
H.~Machner, J.~Haidenbauer: \emph{J. Phys.}, G 25, R231 (1998).

\bibitem{Moskal_Review}
P.~Moskal et~al.: \emph{Progress in Part. Nucl. Phys.}, 49, 1
(2002).

\bibitem{Betigeri02}
M.~Betigeri et~al.: \emph{Phys. Rev.}, C 65, 064001 (2002).

\bibitem{Moskal05}
P.~Moskal et~al.: \emph{arXive}, hep-ex, 0411052 (2005).

\bibitem{PDG}
K.~Hagiwara et~al. (PDG): \emph{Phys. Rev.}, D 66, 010001 (2002).

\bibitem{Lai02}
A.~Lai et~al.: \emph{Phys. Lett.}, B 533, 196 (2002).

\bibitem{Abdel-Bary05}
M.~Abdel-Bary et~al. (GEM~Collaboration): \emph{Phys. Lett.}, B
(submitted),
  xxx (2005).

\bibitem{Betigeri01}
M.~Betigeri et~al.: \emph{Nucl. Phys.}, A 690, 473 (2001).

\bibitem{Abdel-Samad03}
S.~Abdel-Samad et~al.: \emph{Phys. Lett.}, B 553, 32 (2003).

\bibitem{Locher74}
M.~P. Locher, H.~J. Weber: \emph{Nucl. Phys.}, B 76, 400 (1974).

\bibitem{Canton05}
L.~Canton, L.~G. Levchuk: \emph{Phys. Rev.}, C 71, 04041001 (2005).

\bibitem{Willis97}
N.~Willis et~al.: \emph{Phys. Lett.}, B 406, 14 (1997).

\bibitem{Sibirtsev03}
A.~Sibirtsev et~al.: \emph{arXive}, nucl-th, 0310079 (2003).

\bibitem{Khemchandani_02}
K.~P. Khemchandani, N.~G. Kelkar, B.~K. Jain: \emph{Nucl. Phys.},
A708, 312
  (2002).

\bibitem{Betigeri00}
M.~Betigeri et~al.: \emph{Phys. Lett.}, B 472, 267 (2000).

\bibitem{Banaigs73}
J.~Banaigs, J.~Berger, L.~Goldzahl, T.~Risser, L.~Vu-Hai,
M.~Cottereau, C.~Le
  Brun: \emph{Phys. Lett.}, 45B, 394 (1973).

\bibitem{Kirchner93}
T.~Kirchner: Dissertation, Inst. de Physique Nucleaire (Orsay)
(1993).

\bibitem{Bilger02}
R.~Bilger et~al.: \emph{Phys. Rev.}, C 65, 044608 (2002).

\bibitem{Stenmark03}
M.~Stenmark: \emph{Phys. Rev.}, C 67, 034906 (2003).

\bibitem{Khemchandani_03}
K.~P. Khemchandani~N. G., Kelkar, B.~K. Jain: \emph{Phys. Rev.},
C68, 064610
  (2003).

\bibitem{Magiera00}
A.~Magiera, H.~Machner: \emph{Nucl. Phys.}, A 674, 515 (2000).

\bibitem{Abdel-Bary03}
M.~Abdel-Bary et~al.: \emph{Phys. Rev.}, C 68, 021603 R (2003).

\bibitem{Baru03}
V.~Baru, J.~Haidenbauer, C.~Hanhart, J.~A. Niskanen: \emph{Phys.
Rev.}, C 68,
  35203 (2003).

\bibitem{WASA}
WASA at~COSY: www.fz-juelich.de/ikp/wasa/data (2004).

\end{thebibliography}

   \end{document}